УДК 004.9:66.013.512

# ОСОБЕННОСТИ КОМПЛЕКСНОЙ САПР РЕКОНСТРУКЦИИ ПРОМЫШЛЕННЫХ ПРЕДПРИЯТИЙ

## В.В. Мигунов[1]

Обсуждаются особенности разработки проектов реконструкции действующего предприятия силами его проектно-конструкторского подразделения: конечным результатом являются чертежи по стандартам ЕСКД и СПДС; большое число малых проектов для различных имеющихся объектов; разнообразие марок чертежей в составе проекта; большой бумажный архив. На примере САПР TechnoCAD GlassX изложены модели и методы разработки комплексной САПР с удобной единой средой проектирования, с настройкой на профиль работ, с использованием общих частей проекта при наличии ряда проблемно-ориентированных подсистем.

## Введение. САПР в реконструкции предприятий

Проекты реконструкции промышленных предприятий охватывают много небольших объектов с различных сторон – технологической, монтажной, строительной, электротехнической, санитарно-технической и др. согласно требованиям системы проектной документации для строительства (СПДС). В этих условиях для целей автоматизации приходится набирать несколько различных САПР [Орельяна, 2001] с естественными сложностями в освоении их интерфейса и неудобством работы в разных средах. Обычно такие проекты выполняются силами небольших проектно-конструкторских подразделений (ПКО) самого предприятия, что делает невозможной специализацию рабочих мест и усиливает эти трудности. В чертежах есть общие части: строительная подоснова, технологическая схема. Они должны передаваться от одного проектировщика к другому и допускать доработку на всех рабочих местах.

Результатом работы ПКО является проектно-сметная документация для строительно-монтажных работ (СМР), в основном чертежи различных марок СПДС. Электронное представление чертежей не дает значимого эффекта при выполнении почти не автоматизируемых СМР, и эффективность САПР реконструкции не распространяется на их производство. В самом ПКО, как правило, большая часть чертежей хранится в бумажном архиве, и лишь по мере проектирования реконструкции переводится в электронную форму, что также снижает эффективность применения САПР. В результате наблюдается отставание в развитии САПР реконструкции предприятий от САПР в других сферах.

---

[1] *420088, Казань, ул. Губкина, 50, ЦЭСИ Р при КМ РТ, vtigunov@csp.kazan.ru*

Задача автоматизации проектирования реконструкции промышленного предприятия требует программного обеспечения, ориентированного на выпуск чертежей по требованиям ЕСКД и СПДС, работающего одновременно с растровой и векторной графикой, в едином интерфейсе пользователя автоматизирующего подготовку чертежей различных марок СПДС в едином графическом формате. Теории таких систем практически нет. Из 692 диссертаций по специальности 05.13.12 (САПР), защищенных в период 1988 – 2002 г.г. с передачей авторефератов в РГБ, ни одна не посвящена этой проблеме (выборка с http://aleph.rsl.ru, 10.02.2004).

Ниже на примере отечественной САПР TechnoCAD GlassX [Мигунов, 2004] изложены некоторые модели и методы разработки комплексной САПР, обеспечивающей удобную единую среду проектирования с настройкой на профиль работ, с использованием общих частей проекта при наличии ряда проблемно-ориентированных подсистем. Состав подсистем сформирован на основе практических потребностей действующего ПКО ОАО "Казаньоргсинтез" в период 1994 – 2004 г.г.

## Модель чертежа

Объектом разработки является двумерный чертеж – совокупность плоских геометрических элементов. Временный выход из плоских представлений допускается в специализированных расширениях (СР), таких как проектирование аксонометрических схем трубопроводных систем (АСТС) или молниезащиты зданий и сооружений (МЗС). Габариты чертежа достигают 30м для технологических схем. Элементы с размером на бумаге менее 0.3 мм, трудно различимые невооруженным взглядом, не принимаются в чертеж. Применяются две системы координат: натурная и бумажная, с единицами измерения мм, с началом в середине чертежа. Координаты хранятся в четырехбайтных числах: при 23 битах в мантиссе различимы точки 15000мм и 15000.001мм. На время геометрических построений эти числа преобразуются в рабочий 8-байтный формат, что при 52 битах мантиссы повышает разрешение еще в 500 млн. раз. В есть элементы, хранящие вместе с геометрической частью комплекты параметров для реализации специализированных операций (модуль). Обычно один файл чертежа содержит один лист (одну основную надпись).

## Работа с растрами

Значительная часть чертежей выполняется "поверх" прежних чертежей реконструируемых объектов внесением в них фрагментарных изменений. Неизменяемая часть чертежа остается в растровом виде, нужные для стыковки с реконструируемой частью элементы обводятся на растровой подложке, а новая часть вычерчивается сразу в векторном виде. Для обводки растра применяются средства подгонки по задаваемому приближению геометрического элемента. В чертеже одновременно присутствуют и выводятся на печать и векторная, и растровая части. Достаточно работать с монохромными растрами с разрешением 75 – 300 точек на дюйм, что обеспечивает нормальное качество при печати. Используется склейка растров. Вывод растра на экран производится по частям, без ожидания вывода частей, находящихся за экраном.

## Выполнение требований ЕСКД и СПДС

Использование иностранных графических ядер приводит к необходимости в специальных надстройках для выполнения требований ЕСКД и СПДС, таких, как auto.ЕСКД и auto.СПДС [РПК, 2004]. Естественно выполнять эти требования не факультативно, а в той или иной степени принудительно: за счет специальной модели чертежа, не допускающей нестандартных вариантов (жесткое принуждение), и за счет большего удобства применения стандартных решений (мягкое принуждение). Жестко ограничиваются выбираемые варианты: типов линий, масштабов, размера и угла наклона чертежного шрифта, величины стрелок, засечек, перехода выносных линий за размерные, проекций, бланков табличных конструкторских документов (ТКД). Мягкое принуждение реализовано в выборе формата чертежа (пользовательский задавать дольше); в штриховании (пользовательский стиль сделать сложнее); в графических библиотеках: в комплекте GlassX содержатся стандартные основные и дополнительные надписи, условные обозначения арматуры, элементов трубопроводов и др. (пользовательские библиотеки создавать сложнее, чем выбрать из имеющихся); специальный тип геометрического элемента "Магистраль" упрощает вычерчивание условных обозначений трасс кабелей, проводок и др.; все СР ускоряют проектирование, но позволяют создавать чертеж только по стандартам.

## Электронные номенклатурные каталоги

Электронные каталоги изделий, выпускаемых промышленностью (ЭНК) включают сведения из ГОСТов, ОСТов, ТУ, каталогов производителей оборудования, труб, арматуры, элементов трубопроводов, приборов и исполнительных механизмов. Они охватывают (пока) задачи специфицирования по трем профилям работ для чертежей марок: МТ (монтажно-технологическая часть проекта, охват 111 бумажных каталогов), А.. (автоматизация, 14 каталогов); ОВ, ВК, НВК, ТС (отопление, вентиляция, водоснабжение, канализация, теплоснабжение, 102 каталога). Все каталоги для целей САПР удалось представить в единой форме таблиц с данными и текстовых правил генерации обозначения, наименования и других полей спецификаций в едином пользовательском интерфейсе. Графическая часть бумажных каталогов в ЭНК не является необходимой. В данные в таблицах можно помещать списки допустимых вариантов (встроенные меню), а в правила - внешние меню и формы ввода. Правила задают сложение фрагментов строк, которые могут быть константой, значением поля в таблице данных, результатом выбора из внешнего меню, результатом ввода числового или строкового значения, а также временной переменной, в которой сохранено введенное, выбранное или сгенерированное ранее для другой графы ТКД значение. Правила и встроенные меню запускаются в работу, когда проектировщик выбрал подходящее изделие в таблице данных. В ТКД числовые значения приводятся к нужным единицам измерения автоматически, а в графах таблиц данных сохраняются единицы измерения из бумажных каталогов. Для ускорения выбора из 68213 строк 265 таблиц имеются интерактивные и автоматические фильтры по профилю работ, по исходному каталогу, по символу измеряемой прибором величины, по признаку первичный – вторичный прибор, по принадлежности изделий группам в иерархической классификации ("Оборудование/Насос"), по интервалам температур, давлений и др., по значениям в любой графе таблицы данных.

## Специфицирование

Модель ТКД для автоматизации специфицирования с использованием ЭНК реализуется путем совместного хранения в элементе чертежа типа "Табличный модуль" видимых геометрических элементов и параметрического представления (ПП) ТКД, достаточного для генерации его изображения. Первично ПП в виде массива записей. Нулевая запись – шапка таблицы. Структура записей одинакова и задается иерархией деления блоков (прямоугольных фрагментов ТКД) на части по вертикали и горизонтали, с признаками видимости этого деления в шапке и в области данных, на фиксированное или произвольное число частей, вплоть до неделимых блоков – ячеек таблицы. Таким образом задаются ТКД с разделами или без них. Видимые в шапке ячейки имеют свойство "текст в шапке" – заголовки граф. У блока может быть ключевое слово для поиска в ЭНК ("Трубы" – выбор в ЭНК труб). Ячейкам можно задать единицы измерения. В ТКД заносятся пересчитанные значения, например, давления в МПа, хотя в ЭНК встречаются и кгс/кв.см, и м вод.ст.

Генерация спецификаций, заказных спецификаций, таблиц колодцев, монтажных ведомостей трубопроводов и др. достигается путем хранения невидимых специфицирующих свойств в модулях чертежа: АСТС, приборы и исполнительные механизмы, профили наружных сетей водоснабжения и канализации (ПНС), позиционные обозначения. Задается область сбора информации (текущий чертеж или несколько чертежей на диске, типы модулей), и производится автоматическое заполнение ТКД. Для выделения разделов и оформления ТКД имеются автоматизированные операции фасовки строк, упорядочивания по задаваемому списку граф, сливания одинаковых и выделения общих частей наименований. "Буфер изделий" позволяет переносить группу строк с одноименными специфицирующими свойствами между различными ТКД. Можно продолжать ТКД влево или вправо кусками нужной высоты, повторяя шапку или строки с номерами граф, менять типы линий, шрифт текста в графе или ячейке и др. с мгновенной перегенерацией изображения.

Модульная технология создания специализированных расширений

СР эффективны, когда трудоемкость ввода данных в СР существенно ниже трудоемкости непосредственного черчения и расчетов. Например, когда нужны трудоемкие расчеты или нормативные требования порождают много графических изображений по малому объему данных. В обоих случаях наиболее эффективна генерация чертежа по параметрическому представлению. Каждое СР реализуется в своем основном меню как группа операций, в ходе которых создается ПП объекта проектирования. На экране постоянно поддерживается изображение, соответствующее текущему ПП. Затем геометрическая часть и ПП помещаются в чертеж как модуль соответствующего типа. Как элемент чертежа, модуль может выбираться, сдвигаться, удаляться, помещаться в графическую библиотеку, к нему возможны привязки и др.

Тип модуля задает структуру его ПП: списки объектов и общие параметры, часть из которых носят смысл установок. Списки объектов – массивы, за каждым их элементом закрепляются все сведения об объекте, включающие в зависимости от его сути геометрические характеристики, признаки ориентации, специфицирующие свойства, цвет, тип линии, тексты, ссылки на объекты других списков по номерам и др. Например, тексты обычно привязаны к другим элементам чертежа, и списки текстов имеют ссылки на них. Совокупность списков объектов – реляционная база данных со ссылочной целостностью. Специализированные структуры данных в ПП сильно повышают возможную степень автоматизации работ. Хранение ПП на диске создает удобные библиотеки прототипов.

На основе модульной технологии созданы СР по проектированию: строительной подосновы, схем автоматизации, АСТС, ПНС, ТКД, МЗС. Наиболее критично выделение части проекта, подготовка которой будет автоматизироваться в СР. С одной стороны, желательно охватить широкий круг задач. С другой стороны, задачи должны допускать общее внутреннее представление. В некоторых случаях индивидуальность задачи столь сильна, что выбор части проекта становится тривиальным (МЗС). В то же время АСТС очень близки в таких частях проекта, как чертежи технологических трубопроводов, схемы водопровода и канализации, отопления, теплоснабжения, вентиляции, кондиционирования воздуха.

## Интерфейс с пользователем

Состав опций главного меню определяется выбором текущего профиля работ из 5 вариантов: "Монтажно-технологический", "КИП и автоматика", "Строительный", "Электротехнический", "Производственные регламенты" (не для ПКО). В соответствии с этим профилем доступны различные наборы операций, комплекты таблиц ЭНК, меняются свойства части меню и форм ввода, установки по умолчанию и др. Контекстная справка по текущему режиму работы всегда доступна, содержит 12 тыс. строк в 1300 разделах. Текст справки начинается с содержательной характеристики ситуации и заканчивается описанием правил работы в текущем состоянии пользовательского интерфейса (для начинающих пользователей).

## Заключение

Изложенные модели и методы разработки комплексной САПР реконструкции промышленных предприятий не охватывают всего разнообразия чертежей в проектах реконструкции, но являются элементами технологии разработки САПР для них, прошедшими апробацию на практике в ходе десятилетнего развития системы TechnoCAD GlassX. Подробнее о ней: http://technocad.narod.ru.

## Список литературы